\documentclass[aps,prb,reprint,twocolumn,superscriptaddress]{revtex4-1}
\usepackage{amsmath}
\usepackage{amsfonts}
\usepackage{graphicx}
\usepackage{dcolumn}
\usepackage{bm}
\usepackage{amsmath}
\usepackage{hyperref}
\usepackage{float}
\usepackage{color}
\usepackage{natbib}
\usepackage{multirow}

\begin{document}
\title{Unidimensional model of the ad-atom diffusion on a substrate submitted to a standing acoustic wave II. Solutions of the ad-atom motion equation}
\author{C. Taillan}
\affiliation{CNRS; CEMES (Centre d'Elaboration des Mat\'eriaux et d'Etudes Structurales); BP 94347, 29  rue J. Marvig, F-31055 Toulouse, France.}
\affiliation{Universit\'e de Toulouse; UPS; F-31055 Toulouse, France}

\author{N. Combe}
\affiliation{CNRS; CEMES (Centre d'Elaboration des Mat\'eriaux et d'Etudes Structurales); BP 94347, 29  rue J. Marvig, F-31055 Toulouse, France.}
\affiliation{Universit\'e de Toulouse; UPS; F-31055 Toulouse, France}

\author{J. Morillo}
\affiliation{CNRS; CEMES (Centre d'Elaboration des Mat\'eriaux et d'Etudes Structurales); BP 94347, 29  rue J. Marvig, F-31055 Toulouse, France.}
\affiliation{Universit\'e de Toulouse; UPS; F-31055 Toulouse, France}
\begin{abstract}
 The ad-atom dynamic equation, a Langevin type equation is analyzed and solved using some non-linear analytical and numerical tools.  We noticeably show that the effect of the surface acoustic wave is to induce an effective potential that governs the diffusion of the ad-atom: the minima of this effective potential correspond to the preferential sites in which the ad-atom spends more time.  The strength of this effective potential is compared to the destructuring role of the thermal diffusion and to the crystalline potential induced by the substrate.
\end{abstract}

\maketitle
\section{Introduction} 
\label{sec:intro}

The self-organization of materials at the nanoscale is a promising way to avoid the expensive lithography stage in the conception of  the semi-conductors devices. Common self-assembling techniques rely on the Stranski-Krastanov growth modes,~\cite{pimpinelli_villain,Ross1998} buried dislocations networks,~\cite{Brune1998,Leroy2005} or nano-patterned substrates.~\cite{zhong2004,turala2009,jin1999,Mohan2010}
An alternative approach to self-assemble materials at the nano-scale, \textit{the dynamic substrate structuring effect} has been recently proposed.~\cite{Taillan2011} In this approach, a Standing Acoustic Wave (StAW) governs the diffusion of the ad-atoms on a substrate.  In the first paper of this series,~\cite{Combe2011} we have established an unidimensional model of an ad-atom diffusing on a substrate submitted to a StAW and derived the ad-atom motion equation, a Generalized Langevin equation:
 \begin{eqnarray} 
 m \frac{d^2x}{dt^2}&+& \int_{t_0}^{t} \gamma(x(t),x(t'),t-t') \frac{dx}{dt'}(t') dt'  =\nonumber \\
  & & - \frac{d\Phi_{\rm eff}}{dx}(x) +  \xi(t) + F_{SAW}(x,t).  \label{eq:total}
 \end{eqnarray} 
The first left hand side (l.h.s.) term of Eq.~\eqref{eq:total} is the usual inertial term. The second l.h.s. is a retarded friction force with memory kernel $\gamma(x(t),x(t'),t-t')$ where $t_0$ corresponds to the time when the StAW production mechanism is switched on: since $\gamma(x(t),x(t'),t-t')$ is a decaying function of $|t-t'|$, we can fix $t_0$ at $-\infty$ in the integral without loss of generality.
The first right hand side (r.h.s.) term  is an effective inter-atomic (substrate-adatom) periodic potential force derived from the effective potential $\Phi_{\rm eff}(x)$, the second r.h.s. term $\xi(t)$ is a stochastic force and the last one $F_{SAW}(x,t)$ is an effective force induced by the StAW.~\cite{Combe2011} 

The goal of this paper is to study the solutions of Eq.~\eqref{eq:total} and to evidence the structuring role of the StAW on the adatom diffusion, through the effective  $F_{SAW}(x,t)$ force. 
In  Ref.~\onlinecite{Combe2011} we showed that  $F_{SAW}(x,t)$  reads $F_{SAW} \cos(k x+\varphi) \cos (\omega t)$ with $k$ and $\omega$ the wave vector and angular frequency of the StAW.  
However, due to the precise nature of the ad-atom-substrate interactions, $F_{SAW}(x,t)$ experiences also some variation at the substrate lattice parameter scale:  the proportionality factor $F_{SAW}$ and the phase factor $\varphi$ vary as a function of the exact position of the ad-atom in between two successive atomic substrate potential wells.
Because we focus here on the structuring role of the StAW, we will only consider its large scale variation, i.e. $F_{SAW}(x,t)= F_{SAW} \sin(k x) \cos (\omega t)$ with a constant $F_{SAW}$ value and a constant phase factor that we fix at $-\pi/2$ for convenience, eluding thus the possibility for this force to vary on the substrate lattice parameter scale: such dependence essentially affects the dynamics of the ad-atom  and weakly the structuring effects of the StAW.\cite{Taillan2011} The detailed study of the dynamics of the ad-atom will be reported elsewhere.

Concerning the stochastic force $\xi(t)$ and the memory kernel $\gamma(x,x',t-t')$, which are related by the fluctuation-dissipation theorem:\cite{Zwanzig1973,Combe2011} 
\begin{equation}
\langle \xi(t) \xi(t+\tau)\rangle = k_BT \gamma(x(t),x(t+\tau),\tau),\label{eq:flucdis}
\end{equation}
with $k_B$ the Boltzmann constant and $T$ the temperature, their properties are very interaction-potential model dependant and thus cannot be studied on a general ground. 
Though we have established their analytical expressions in a very peculiar potential case in the preceeding paper,~\cite{Combe2011} we will here use the more standard and general model of a centered gaussian noise for the stochastic force, with correlation time $\tau_c$.~\cite{Pottier2007,Gordon2008,Forster1990}

\begin{itemize}
\item The stochastic force is then fully characterized by its zero mean value and autocorrelation function: 
\begin{subequations}\label{eq:corr-tau}
\begin{equation}
\langle\xi(t)\rangle=0, \label{mean}
\end{equation}
\begin{equation}
\langle \xi(t) \xi(t+\tau)\rangle =  D \frac{e^{-|\tau|/\tau_c}}{\tau_c}; \label{eq:corr}
\end{equation}
\end{subequations}
\item and the memory kernel is given by (Eq.~\eqref{eq:flucdis}):
\begin{equation}
\gamma(x(t),x(t+\tau),\tau) = \frac{\gamma}{\tau_c} e^{\frac{-|\tau|}{\tau_c} },\label{eq:mem_kernel}
\end{equation} 
\end{itemize}
where $D/m^2$ is the adatom diffusion coefficient in the velocity space in the absence of the StAW and $\gamma$ is the friction coefficient, given by $ \gamma= k_B T/D $.
Eqs.~\eqref{eq:corr-tau} relies on the high number of atoms in the substrate and its thermalization by an external macroscopic system in experimental conditions.~\cite{Pottier2007,Combe2011} 
In this model, the correlation time $\tau_c$ is of the order of the inverse of the Debye frequency of the substrate. The meaning of the friction coefficient appears clearly in the limit of vanishing correlation time : 
\begin{subequations}\label{eq:corr-tau1}
\begin{equation}
\langle \xi(t) \xi(t+\tau)\rangle \xrightarrow{ \tau_c \to0}  2 D \delta(\tau), \label{eq:corr-tau0}
\end{equation}
where $\delta$ is the Dirac distribution, and thus
\begin{equation}
 \int_{-\infty}^{t} \gamma(x(t),x(t),t-t') \frac{dx}{dt'}(t') dt'   \xrightarrow{ \tau_c  \to 0}  \gamma    \frac{dx}{dt}(t),   
\end{equation}
the usual friction force proportional to the speed.
In this limit, $m/\gamma$ is  the relaxation time $\tau_R$ of the ad-atom dynamics in the absence of r.h.s. in Eq.~\eqref{eq:total}.
\end{subequations}

Eq.~\eqref{eq:total} involves two different length-scales, the StAW wave-length $\lambda= 2 \pi/k$ and the periodicity $a$ of the crystalline potential; and three different time-scales, the correlation time $\tau_c$, the StAW period $T=2 \pi/\omega$ and the relaxation time $\tau_R$ of the ad-atom dynamics.  In the following analytical calculations, we assume that all these scales are separable and have the following properties:
\begin{subequations}
\begin{align}
  &a  \ll   \lambda& \label{eq:length_scale} \\
  &\tau_c \ll T \ll \tau_R& \label{eq:timescale}
\end{align}\end{subequations}
Eq.~\eqref{eq:length_scale} derives from the fact that we consider StAW wavelengths varying from few to hundreds of  nanometers compared to the substrate lattice parameter of the order of $10^{-1}$ nm. 
Eq.~\eqref{eq:timescale} is motivated by ours Molecular Dynamic (MD) simulations of ad-atom diffusion on a substrate submitted to a nanometer wavelength StAW.~\cite{Taillan2011}  

The effective crystalline potential ($\Phi_{\rm eff}$) will be modelled by its fundamental Fourier component:
\begin{equation}
\Phi_{\rm eff}(x)=- \frac{a\varphi_0}{2\pi} \cos(2\pi x/a), \label{eq:pot0}
\end{equation}
where $\varphi_0$ is the amplitude of the corresponding force. 
Because of the difference of length scales (Eq.~\eqref{eq:length_scale}), we assume for simplicity that the wavelength of the StAW force is an integer multiple of the lattice parameter $\lambda=na$. 
\begin{equation}
\Phi_{\rm eff}(x)=- \frac{\varphi_0}{nk} \cos(nkx). \label{eq:pot}
\end{equation}
Within the presented model, Eq.~\eqref{eq:total} writes now:
 \begin{eqnarray} 
 m \frac{d^2x}{dt^2}+ \frac{\gamma}{\tau_c}\int_{-\infty}^{t}  e^{\frac{-|t-t'|}{\tau_c} } \frac{dx}{dt'}(t') dt'  =\nonumber  \\
    -\varphi_0\sin(nkx) +  \xi(t) + F_{SAW}\sin(k x)& \cos (\omega t),   \label{eq:total1}
 \end{eqnarray} 
where $\xi$ is the Gaussian noise defined by Eqs.~\eqref{eq:corr-tau}. 

In a reduced dimensionless formulation Eq.~\eqref{eq:total1} writes:
 \begin{eqnarray} 
\ddot{\tilde x}&+& \frac{\tilde {\gamma}}{\tilde \tau_c} \int_{-\infty}^{\tilde t} e^{\frac{-|\tilde t-\tilde t'|}{\tilde \tau_c} } \dot{\tilde x} d\tilde t'  =\nonumber \\
  & & -\tilde \varphi_0\sin(n\tilde x)  +  \tilde \xi(\tilde t) + \tilde F \sin(\tilde x) \cos (\tilde t),  \label{eq:totalreduced}
 \end{eqnarray} 
with $\tilde x= kx$, $\tilde t = \omega t$, $\tilde{\gamma}=\gamma/(m\omega)$, $\tilde{\tau}_c = \omega \tau_c$, $\tilde \xi(\tilde t)=k \xi(t)/(m\omega^2)$ and $\tilde{F}=k F_{SAW}/(m\omega^2)$ and where $\dot{\tilde x}$ and $\ddot{\tilde x}$ are the first and second derivatives of $\tilde x$ relative to $\tilde t$ respectively.

Getting an explicit expression of the general solution of the non-linear stochastic integro-differential equation \eqref{eq:total1} or \eqref{eq:totalreduced} is unreachable at least for us and, anyways is not our goal. 
As stated before, what we are interested in, is to evidence the conditions for the appearance of a structuring effect on the ad-atom diffusion due to the effective StAW force and how the other forces can affect it.
 
In Sect.~\ref{sec:maineffect}, we evidence the structuring effect of the StAW force, by considering  Eq.~\eqref{eq:totalreduced} in the long time ($t\gg \tau_c$) and length scales  ($x\gg a$) limit without thermal fluctuations. In this limit, the periodic potential and the stochastic forces can be neglected and the retarded effects in the friction force too (limit $\tau_c \to 0$).  Eq.~\eqref{eq:totalreduced} reduces then to:
 \begin{equation} 
\ddot{\tilde x}+ \tilde {\gamma} \dot{\tilde x}  =
  \tilde F \sin(\tilde x) \cos (\tilde t).  \label{eq:maineffect}
 \end{equation} 
Since Eq.~\eqref{eq:maineffect} is non-linear, we apply, in Sect.~\ref{sec:maineffect} some standard tools of the non-linear physics to characterize its solutions. First, using the multiple time scale analysis,~\cite{Kevorkian1996} we will evidence the existence of an effective potential $U_{\rm eff}$ governing the diffusion in the long time scale limit. The general solutions will then be studied using the fixed points stability analysis. Then, it will be numerically integrated and its solutions will be analyzed with the use of the Poincar\'e sections of the phase diagram and the calculation of their Lyapounov exponents.
The stochastic force $\xi(t)$ will then be reintroduced in Sect.~\ref{sec:noise} in the same $\tau_c \to 0$ limit. It will be shown that it mainly induces some fluctuations around the solutions of Eq.~\eqref{eq:maineffect}. 
In Sect.~\ref{sec:retarded}, the retarded effects  ($\tau_c \neq 0$) will be added, whereas the effects of the substrate effective cristalline potential will be reintroduced in  Sect.~\ref{sec:crystal} in the absence of retarded effects (limit $\tau_c \to 0$). 
In all the sections from~\ref{sec:noise} to~\ref{sec:crystal} , the structuring effect of the StAW force and its sensitivity to the other forces will be evidenced through the study of the position probability density of the ad-atom.
Finally, in Sect.~\ref{sec:pot_eff_distrib} the relevance of the analysis of Sect.~\ref{sec:maineffect} on the solutions of the complete equation (Eq.~\eqref{eq:totalreduced}) will be demonstrated.

\section{Main Effects of the standing acoustic wave on ad-atom diffusion}
\label{sec:maineffect}
We study in this section the non-linear equation Eq.~\eqref{eq:maineffect}. Note that this equation is invariant under a simultaneous space and time translation:  $\tilde x \to \tilde x + \pi$ and $ \tilde t \to\tilde  t+ \pi$. As a consequence,  all the results concerning the structuring effects will be invariant under a space translation  $\tilde x \to\tilde  x + \pi$.

\subsection{Multiple time scale analysis}
\label{sec:multiple}

Eq.~\eqref{eq:maineffect} is a non-linear deterministic equation that implies two different typical time scales: the dimensionless relaxation time  $\tilde{\tau}_R = \omega \tau_R$ (depending on $\tilde{\gamma}$) of the ad-atom dynamics and the period of the StAW force. To manage these time scales, this equation is first analyzed using the multiple scale method.\cite{Kevorkian1996}  This method, due to the extra degrees of freedom it introduces, allows to remove the secular divergencies that can arise in a standard perturbation approach. Note that Eq.~\eqref{eq:maineffect} without friction has already been studied~\cite{Escande1982} in a different framework.

We consider the limit $\epsilon=1/\omega \to 0$ (keeping constant $c_s=\omega/k$, the sound speed in the substrate).  
Eq.~\eqref{eq:maineffect} reads then: 
\begin{equation}
\ddot{\tilde x}+\epsilon \underline{\gamma} \dot{\tilde x} = \epsilon \underline{F} \sin(\tilde x) \cos(\tilde{t}),\label{min}
\end{equation}
where $\underline{\gamma}=\gamma/m$ and $\underline{F}=F_{SAW}/(mc_s)$ are  order 0  quantities ($O(\epsilon^0)$) .
We seek an approximate solution of Eq.~\eqref{min} using the following expansion:
\begin{eqnarray} 
\tilde x(\tilde{t},\epsilon) &=& \tilde x_0(\tilde t_0,\tilde t_1)+\epsilon \tilde x_1(\tilde t_0,\tilde t_1) \nonumber\\
&&+\epsilon^2 \tilde x_2(\tilde t_0,\tilde t_1)+ O(\epsilon^3),\label{me}
\end{eqnarray}
involving the two time scales: $\tilde t_0=\tilde{t}$ and $\tilde t_1=\epsilon \tilde{t}$.
Substituting Eq.~\eqref{me} into Eq.~\eqref{min} and identifying terms of the same order in $\epsilon$, we get the following equations:

\begin{subequations}
\begin{align}
 D_0^2 \tilde x_0 &=& 0, &\label{order0}\\
 D_0^2 \tilde x_1 &=& -&2 D_0 D_1 \tilde x_0 - \underline{\gamma} D_0 \tilde x_0 + \underline{F} \cos(\tilde t_0) \sin(\tilde x_0),\label{order1}\\
 D_0^2 \tilde x_2 &=& - &2 D_0 D_1 \tilde x_1 - D_1^2 \tilde x_0 - \underline{\gamma} ( D_0 \tilde x_1 +  D_1 \tilde x_0 ) \nonumber\\ 
 &&+ & \underline{F} \tilde x_1 \cos(\tilde t_0) \cos( \tilde x_0),\label{order2}
\end{align}
\end{subequations}

where the operator $D_n$ designs $\partial/\partial \tilde t_n$ with $n \in \{0,1\}$.
The solution of Eq.~\eqref{order0} reads:
\begin{equation}
\tilde x_0=A(\tilde t_1) \tilde t_0 + B(\tilde t_1).\label{sol00}
\end{equation}
The first term of Eq.~(\ref{sol00}), a secular term that diverges with $\tilde t_0$,
 is removed by setting  $A(\tilde t_1)=0$, leading to:
\begin{equation}
\tilde x_0=\tilde x_0(\tilde t_1).\label{sol0}
\end{equation}

Substituting Eq.~\eqref{sol0} into Eq.~\eqref{order1},  the particular solution $\tilde x_1$ writes: 
\begin{equation}
\tilde x_1(\tilde t_0,\tilde t_1)= - \underline{F} \cos(\tilde t_0) \sin(\tilde x_0). \label{x1}
\end{equation}
Using these expressions of $\tilde x_0$ (Eq.~\eqref{sol0}) and $\tilde x_1$ (Eq.~\eqref{x1}) in Eq.~\eqref{order2},  we write the solubility condition of this equation
by the elimination of the secular term:
\begin{equation}
D_1^2 \tilde x_0 + \underline{\gamma} D_1 \tilde x_0 = -\frac{\underline{F}^2}{2} \cos( \tilde x_0) \sin( \tilde x_0),\label{secular}
\end{equation}
a differential equation governing the solution $\tilde x_0$ on the time scale $\tilde t_1$.
To give a physical meaning to Eq.~\eqref{secular}, we note that using Eqs.~\eqref{sol0} and  ~\eqref{x1} in Eq.~\eqref{me}, the solution of Eq.~\eqref{eq:maineffect} writes to $O(\epsilon^1)$ order: 
\begin{equation}
\tilde x(\tilde{t})  = \tilde x_0(\tilde t_1) - \tilde{F} \cos(\tilde t_0) \sin (\tilde x_0(\tilde t_1)).
\end{equation}
Therefore, $X(\tilde t_1)$ the average value of  $\tilde x(\tilde{t})$ over a StAW period, writes to $O(\epsilon^1)$ order: 
\begin{equation}
X(\tilde t_1)=\langle\tilde x(\tilde{t})\rangle = \frac{1}{2 \pi} \int_{\tilde t_0}^{\tilde t_0+2\pi} \tilde x(\tilde{t}) d\tilde{t} \approx \tilde x_0(\tilde t_1).
\end{equation}
 
 Substituting $\tilde x_0(\tilde t_1)$ by  $X(\tilde t_1)$ in Eq.~\eqref{secular} an going back to the $\tilde t$ variable, we obtain the  long time evolution equation of $X(\tilde{ t})=\langle\tilde x(\tilde{ t})\rangle$, the mean value of $\tilde x$ over a StAW period:
\begin{equation}
 \frac{d^2X(\tilde{ t})}{d\tilde{t}^2}  + \tilde{\gamma}   \frac{dX(\tilde{t})\rangle}{d\tilde{t}} = -\frac{\tilde{F}^2}{4} \sin(2X(\tilde{ t})).\label{secular2}
\end{equation}
Note that the scheme provided by Landau and Lifshift~\cite{landau_meca} that develops the variable $\tilde x(\tilde{t})$ in Eq.~\eqref{eq:maineffect} as a sum of a slowly varying function $X(\tilde{t})$ and a quickly varying function $\zeta(\tilde{t})$ yields a similar result.~\cite{Taillan2011}  

The multiple scale method has allowed us to transform the non-autonomous Eq.~\eqref{eq:maineffect}  into an autonomous equation (Eq.~\eqref{secular2}) on a longer time scale. 
This equation describes the motion of the ad-atom on a mesoscopic time scale, long compared  to the period of the StAW, but small compared to the relaxation time $\tilde{\tau}_R$ of the ad-atom dynamics. The StAW force acting at the mesoscopic time-scale  derives from  the effective potential  $U_{\rm eff}$:
\begin{equation}
U_{\rm eff}(X) =  \frac{\tilde{F}^2}{4}  \sin^2(X).  \label{ueff}
\end{equation}
This potential, is periodic with period $\pi$ and minima at $X=0[\pi]$. Whatever the initial conditions the long time evolution described by Eq.~\eqref{secular2} will be a dampted evolution towards one of the minima of $U_{\rm eff}$.
Within the described approximation (Eq.~\eqref{secular2}), the StAW leads then to a self-organization of the ad-atoms diffusion into a periodic array with period half the period of the StAW.
In the following, $U_{\rm eff}$  will appear to be an essential tool to interpret the ad-atom trajectories and the structuring effect of the StAW. Note that the (stable) fixed points of Eq.~\eqref{secular2} which are the minima of $U_{\rm eff}$,  are also those of  Eq.~\eqref{eq:maineffect}.

These results are approximated results, we must now come back to Eq.~\eqref{eq:maineffect} to test their relevance in the general case.
We will start with the study of the stability of the fixed  points ($(\tilde x,\dot{\tilde x}) = (0 [\pi],0)$) of Eq.~\eqref{eq:maineffect} which, as we will see now, can differ from that of Eq.~\eqref{secular2} for certain values of the parameters of the equation. 
 
\subsection{Fixed points stability}
 \label{ssec:fix_point}

 As already mentionned,  Eq.~\eqref{eq:maineffect} is invariant under simple spatial  and time translations so that all the fixed points $(\tilde x,\dot{\tilde x}) = (0 [\pi],0)$ are equivalent.
We hence reduce our stability analysis to one fixed point: $ (\tilde x,\dot{\tilde x}) = (0,0)$.
The linerarized version  of Eq.~\eqref{eq:maineffect} around this fixed point reads:
\begin{equation} 
\dot V=A(\tilde{t})V(\tilde{t})\label{systeme}
\end{equation}
with $ V = \begin{pmatrix}\tilde x\\ \dot{\tilde x}\end{pmatrix} $ and $ A(\tilde{t}) = \begin{pmatrix} 0  &  1 \\ \tilde{F} \cos(\tilde{t}) & - \tilde{\gamma} \end{pmatrix}$. \\
We define $R_{\tilde t_0}^{\tilde t}$, the propagator of  Eq.~\eqref{systeme}: 
\begin{equation}
 V(\tilde t_0)   \stackrel{R_{\tilde t_0}^{\tilde t}}{\longrightarrow} V(\tilde t) 
 \end{equation} 
Since $A(\tilde{t})=A(\tilde{t}+2 \pi)$ in Eq.~\eqref{systeme}, the Floquet theory provides the stability of the fixed point $ (\tilde x,\dot{\tilde x}) = (0,0)$ of Eq.~\eqref{eq:maineffect} from the eigenvalues of $R_{0}^{2 \pi}$: the fixed point is stable if all the eigenvalues of  $R_{0}^{2 \pi}$ are inside the unit circle of the complex plane. 

\begin{figure}
\begin{center}
\includegraphics[width=0.9\columnwidth]{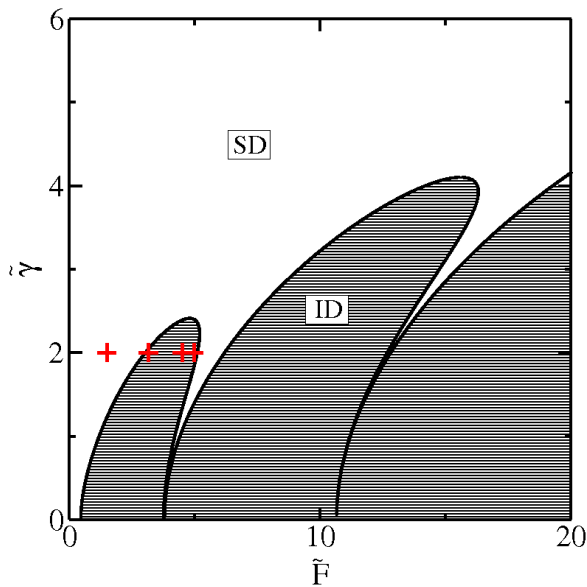}
\caption{Stability (Blank) (SD) and instability (dashed) (ID) domains of the fixed point (0,0) of Eq.~\eqref{eq:maineffect} in the ($\tilde{\gamma}$,$\tilde{F}$) parameter plane (PP).
Red crosses design the couples of parameters chosen in the numerical studies of Sects.~\ref{sec:maineffect},~\ref{sec:noise} and~\ref{sec:retarded}.
 }\label{fig1}
\end{center}
\end{figure}

The propagator $R_{0}^{2 \pi}$ is calculated by numerically integrating Eq.~\eqref{systeme} using a fourth order Runge-Kutta method and is  then diagonalized. Fig.~\ref{fig1} reports the stability diagram of the fixed point $ (\tilde x,\dot{\tilde x}) = (0,0)$ in the  $(\tilde{\gamma},\tilde{F})$ parameter plane (PP).  Fig.~\ref{fig1} reveals the existence of unstable domains which questions the validity of the self-organization effect evidenced for any $(\tilde{\gamma},\tilde{F})$ parameters in the preceding section.
This is not contradictory since the Floquet analysis is exact concerning the fixed points stability, while the approximated multiple scale analysis relies on the smallness of  the amplitude of the fast variations (StAW period time scale) compared to the slow variations ($\tilde{\tau}_R$ time scale) of $\tilde x$.
A condition which will be fulfilled as far as the strength of the friction force ($\tilde \gamma$) remains large enough compared to the strength of the effective StAW force ($\tilde F$), which explains the global separation between the stability (upper left triangle) domain (SD) and the instability (lower right triangle) domain (ID) in the PP.
In addition, in absence of friction $\tilde \gamma=0$, Eq.~\eqref{systeme} is equivalent to a Mathieu equation for a parametric oscillator with a null eigen frequency. The stability diagram of the Mathieu equation can be for instance found in Fig.2 of Ref.~\onlinecite{Ness1967}: it evidences some unstable solution regions, corresponding to the parametric resonances, separated by very tiny stable regions for an oscillator with a null eigen frequency. The large separated unstable subdomains (tongue) in Fig.~\ref{fig1} hence correspond to the parametric resonances that disappear when increasing the dissipation $\tilde \gamma$.


\subsection{Numerical study}
\label{ssec:num_study}

In order to check the self-organisation ability of the effective StAW force in the ID of the PP (Fig.~\ref{fig1}), we need to turn to a precise examination of the trajectories.

An analytical resolution of Eq.~\eqref{eq:maineffect} is out of scope, so we will solve it numerically using the forth order Runge Kutta method for different values of the parameters $(\tilde{\gamma},\tilde{F})$. 
We will also look at the trajectories for parameters in the SD to give a complete scope of the different exact behaviors.
In order to enlighten  some specificities of the trajectories,  the resolution will be performed for different initial conditions $ (\tilde x(0),\dot{\tilde x}(0))$ (IC).
From our analysis of the stability diagram in the preceding section, we expect to observe specific behaviors associated to increasing values of $\tilde F$ at constant $\tilde \gamma$, going from one stability domain to the next one through the midway instability domain.  
This is what we observed in our rather extended exploration of the PP and IC spaces. 
From this exploration, we identified four categories of trajectories: one corresponding to the SD; the three others to the ID, one in its core and the two others in the vicinity of its frontiers with its two neighboring SDs. We did not examine the very peculiar case $\tilde{\gamma}=0$, which corresponds to undamped trajectories,  since it is not relevant for the ad-atom diffusion on a substrate, and since it has been studied previously.~\cite{Escande1982}
Of course,  we cannot absolutely exclude the possibility to have missed some specific behaviors, even if we consider it as highly improbable.
In all the rest of the manuscript, we will focus on the solutions for the  2.0 constant $\tilde \gamma$  value, since it provides a representative sample of the behaviors we have exhibited.
The results are presented for increasing values of $\tilde F$ (1.5, 3.15, 4.5 and 4.96) through the first instability subdomain (red crosses in Fig.~\ref{fig1}). 
In each case, the trajectory is related to the effective potential $U_{\rm eff}(X)$ ( blue solid line in Fig.~\ref{fig2} and Eq.~\eqref{ueff})  in order to evidence any self-organization behavior.

\subsubsection{Converging Trajectories}
\label{sssec:converge}
\begin{figure}
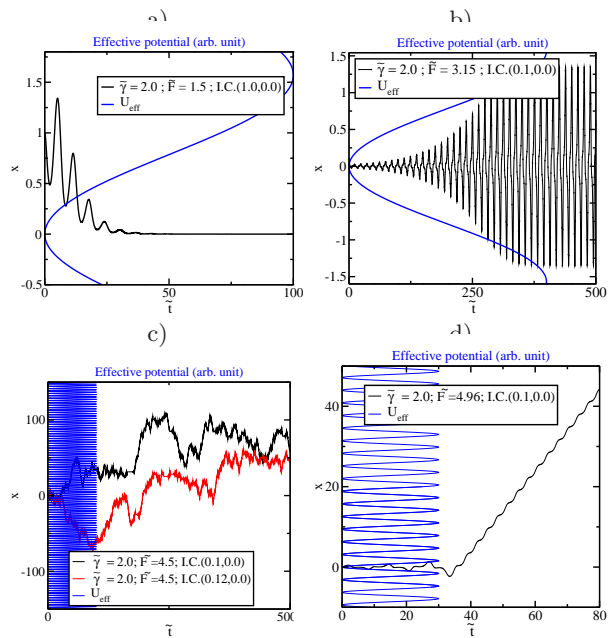

    \centering
    \begin{tabular}{cc}
   a) & b) \\
       \includegraphics[width=.45\linewidth]{fig2a}
     & \includegraphics[width=.45\linewidth]{fig2b}\\
 c) & d) \\
      \includegraphics[width=.45\linewidth]{fig2c} 
     &\includegraphics[width=.45\linewidth]{fig2d}
	\end{tabular}
    \caption{Black and red curves: $\tilde x(\tilde t)$ solutions of Eq.~\eqref{eq:maineffect} for $\tilde{\gamma}=2.0$. Stability domain (SD): a)  converging trajectories ($\tilde{F}=1.5$). Instability domain (ID):  b) periodic trajectories ($\tilde{F}=3.15$), c) chaotic trajectories ($\tilde{F}=4.5$) and d) unbounded trajectories ($\tilde{F}=4.96$). Initial conditions are reported in the legend. Blue curves: $U_{\rm eff}(X)$, the effective potential  (abscissa) in arbitrary units versus position  $X$ (ordinate).}
    \label{fig2}
\end{figure}

The trajectories have typically an oscillating behavior whose amplitude and off-set are decreasing. 
They can be somewhat different at the very beginning, depending on the sign of the $\tilde x(0)\dot{\tilde x}(0)$ product but they have the same character at a longer timescale.
Fig.~\ref{fig2}a reports the solution for initial conditions $(1.0,0.0)$ corresponding to an initial position in the $U_{\rm eff}(X)$ potential valley associated to its 0.0 minimum.
We choose an initial position (1.0) rather away from the minimum of  $U_{\rm eff}(X)$ to best evidence the amplitude and off-set decreases. The initial speed has been fixed to zero in order to only present the characteristic time behavior.

The trajectories converge towards the fixed point  $(0,0)$. 
Depending on the initial conditions, 
 the trajectory can eventually escape from the 0.0 minimum to an adjacent minima. In that case the trajectory will converge to the corresponding $U_{\rm eff}(X)$ minima (fixed point).
The several time scales mentioned in Sect.~\ref{sec:multiple} are  clearly visible: fast oscillations at a $2 \pi$ period (StAW period) whose amplitude slowly decays on the $\tilde{\tau}_R$ time scale.  
The period of the oscillations can be one of the harmonics of the StAW period.
For example, in the vicinity of the stability/instability frontier $(\tilde  \gamma, \tilde F)=(2.0,3.028)$, we observe the second harmonic period ($4 \pi$).
It appears then that, in the SD, the solutions of Eq.~\eqref{eq:maineffect} can be adequately estimated using a multiple time scale analysis, the long time behavior being well described by Eq.~\eqref{secular2}.

\subsubsection{Periodic trajectories}
\label{sssec:periodic} 
In the ID, in the vicinity of the first frontier between the SD and the ID, the unstable character of the fixed point results in an amplification of the oscillations (Fig.~\ref{fig2}b). 
This amplification occurs up to an upper limit fixed by the non-linearities of the sinus in Eq.~\eqref{eq:maineffect}. At this time the trajectory becomes periodic, revealing the existence of a limit cycle, the attractor of the system in the phase space.
The  IC of the presented trajectory  $(\tilde x(0),\dot{\tilde x}(0))=(0.1,0.0)$ correspond to  a position very close to the unstable fixed point  to evidence the amplification of the fast oscillations (transient regime) leading to the periodic trajectory. 
The period of oscillation is $4 \pi$, i.e. twice the excitation period. 
Like in the converging trajectories, this period depends on the peculiar choice of the parameters (see Sect.~\ref{ssec:transitio_to_chaos}). 

The trajectory of Fig.~\ref{fig2}b oscillates in its initial effective potential basin. However, increasing  $\tilde{F}$, results in an increase of the upper limit. This limit may overcome the maxima of $U_{\rm eff}(X)$ so that the trajectory may extend on the two neighboring effective potential basins: however, the average position of the adatom always belong to the same potential basin.
$U_{\rm eff}(X)$, if relevant to characterize the trajectory of the adatom, does not capture all the information contained in the phase space diagram. 
With this kind of trajectories the self-organisation character of the effective StAW force is not apparent but will be demonstrated in Sect.~\ref{sec:noise} studying the distribution of the ad-atom x-position.  


\subsubsection{Chaotic trajectories}
\label{sssec:chaotic}
In order to evidence the chaotic character of some trajectories in the core of the ID region ($\tilde{F}=4.5$) we report on Fig.~\ref{fig2}c the solutions of Eq.~\eqref{eq:maineffect}  for two sets of very closed I.C.s $(0.10,0.0)$ and $(0.12,0.0)$.  Both trajectories visit different effective potential basins following an apparently erratic motion.  
The two solutions, diverge quickly from each other, despite their very close ICs, suggesting the chaotic character of these trajectories. 
To examine this chaotic behavior, we calculate the Lyapounov coefficients ($\lambda_i$) of the autonomous system associated to Eq.~\eqref{eq:maineffect}.  
For the first trajectory, using the algorithm provided by Wolf et al,~\cite{Wolf1985} the highest Lyapounov coefficient $\lambda_L$ is  found to be positive ($0.278$) (binary base), an unambiguous evidence of the chaotic character of this solution.

The transition to the chaotic behavior will be discussed in a separate section (Sect.~\ref{ssec:transitio_to_chaos}). At first sight, it seems that the self-organization character is lost, but as we will see later (Sect.~\ref{sec:noise}) studying the position probability density of the ad-atom, it actually is preserved.


\subsubsection{Unbounded trajectories}
\label{sssec:unbounded}
 We look now, in the ID domain, at a characteristic trajectory in the vicinity of the second frontier between the ID and the SD ($\tilde{F}=4.96$).
The IC of the trajectory reported in Fig.~\ref{fig2}d are the same as those in the two preceding cases, $(0.1,0.0)$. 
After a transient period where the particle stays in its original potential well, it leaves it, without being captured by any other potential well:  the trajectory is unbounded.
As one can see, from this point the trajectory is roughly linear and thus does not present any visible chaotic character. 
This is confirmed by the calculation of its Lyapounov exponents that are all negative (or null). After the transient period, the trajectory is monotonously increasing with a staircase character. Changing the IC can lead to a monotonously decreasing trajectory with a symmetric staircase character.
These trajectories are analogous to the rotations of a pendulum about its pivot in the clockwise or counterclockwise directions and have been already evidenced  in the phase space $(\tilde x(0),\dot{\tilde x}(0))$ in the absence of dissipation $\tilde \gamma = 0$.~\cite{Escande1982}

The final remark of the preceding section still applies here: contrary to the appearances, the self-organization character is preserved.
It infers from the staircase character of the trajectory which corresponds to longer residence times in the potential valleys than the transition times between valleys. 
The self-organization character will be evidenced for all the types of trajectories in Sect.\ref{sec:noise}, where the probability distribution will be studied in the presence of the fluctuating force (vanishing fluctuating force case). 

\subsection{Transition to Chaos}
\label{ssec:transitio_to_chaos}

\begin{figure}
	\includegraphics[width=.9\linewidth]{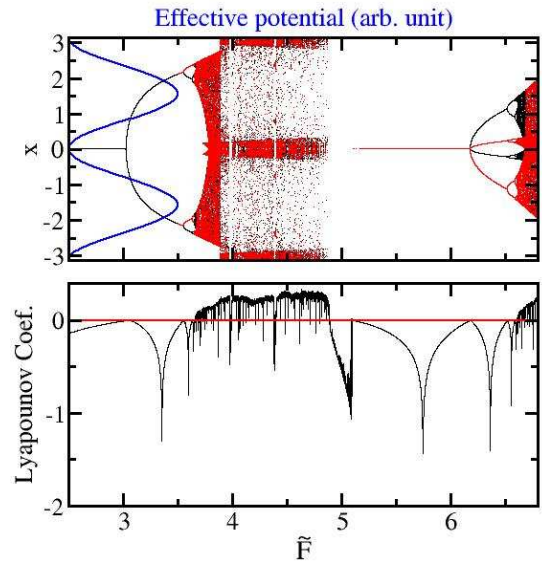}\\
	\caption{a) Poincar\'e section of the phase space omitting transient regime as a function of  $\tilde{F}$  with $\tilde{\gamma}=2.0$ from trajectories calculated by numerical resolution of Eq.~\eqref{eq:maineffect} with initial conditions $(0.1,0.0)$ (black) and $(-0.1,0.0)$ (red).  Blue curve: $U_{\rm eff}(X)$, the effective potential  (abscissa) in arbitrary units versus position  $X$ (ordinate) b) $lambda_L$ the greatest non-null Lyapounov exponent (binary base) of Eq.~\eqref{eq:maineffect} as a function of $\tilde{F}$ for  initial conditions $(0.1,0.0)$.  The solid horizontal line ($lambda_L=0$)  is a guide  to the eyes.}
    \label{fig3}
\end{figure}

We will now characterize the domains of existence, in the PP plane, of the four observed types of trajectories, through the study of the Poincar\'e section ($\dot{\tilde x}=0$) of the solutions of Eq.~\eqref{eq:maineffect} at times larger than the transient initial period.
As in the preceding section, we explore the PP at the constant $\tilde \gamma=2$ value.
We increase $\tilde{F}$ starting with a value $\tilde{F}=1.5$ in the first SD  up to $\tilde{F}=6.84$ in the heart of the second ID, going through the first  ID in between (Fig.~\ref{fig1}). 
The critical values of $\tilde{F}$ separating different behaviors have been calculated using the shooting and continuation methods~\cite{Dooren1996} and are given here with a precision of 0.001. 
Note that these $\tilde{F}$ critical values depend on the $\tilde{\gamma}$ value. The calculations have been performed for a wide range of ICs. Whatever the parameters, we found at most two types of asymptotic trajectories presented in Fig.~\ref{fig3}a (black and red dots) depending on the ICs. The presented points correspond to $(0.1,0.0)$ (black dots) and $(-0.1,0.0)$ (red dots) ICs.
$\lambda_L$, the greatest non-null Lyapounov coefficient (binary base) of the autonomous system associated to Eq.~\eqref{eq:maineffect}, has  been calculated also to characterize the chaotic or not character of the trajectories.\cite{Wolf1985} 
Since the results are independent of the ICs, we present in Fig.~\ref{fig3}b its evolution with $\tilde{F}$  for the $(0.1,0.0)$ IC only.

From Fig.~\ref{fig3} we see that  in the first SD i.e. $\tilde{F}<\tilde{F}_1$, with $\tilde{F}_1=3.028$ the first SD/ID limit, the asymptotic solution of  Eq.~\eqref{eq:maineffect} is the stable fixed point $(0,0)$ whatever the ICs (sec.~\ref{sssec:converge}). This result is coherent with the values of $\lambda_L<0$. 

At $\tilde{F}_1$, $\lambda_L$ goes to 0 for the first time and we enter the first ID, the system undergoes a Hopf bifurcation towards an unique limit cycle (with twice the StAW period),  the attractor of the flow (identical black and red points) in agreement with the results of sec.~\ref{sssec:periodic}.
More precisely, due to the translation invariance of Eq.~\eqref{eq:maineffect}, there is actually an infinite number of limit cycles, attractors of the system, one per fixed point $(0 [\pi],0)$. 
This domain of periodic asymptotic trajectories extends from $\tilde{F}_1$ to $\tilde{F}_2= 3.654$.
Beyond, for $ \tilde F > \tilde{F}_2$,  $\lambda_L$ becomes positive and thus the trajectories chaotic.
The first periodic-chaotic transition is thus at $\tilde{F}=\tilde{F}_2$.

In this periodic domain$ \tilde{F}_1<  \tilde F < \tilde{F}_2$, at $\tilde{F}=3.552$, $\lambda_L$ goes to 0 and the limit cycle splits into two limit cycles (separation of the black and red points) without any period change. The system has then two attractors per fixed point. 
This doubling of the number of attractors, precedes for each attractor a cascade of period doublings at increasingly close values of $\tilde{F}$, each of them being associated to the cancellation of $\lambda_L$. We have been able to observe 4 period doublings (at 3.632,3.650,3.653,3.654), though we did not try to optimize this number.
Each period doubling corresponds to the apparition of the corresponding  peak in the Fourier spectrum.   
The transition to chaos proceeds when the spectrum becomes continuous at $\tilde{F}_2$.

Between $\tilde{F}_2$ and $\tilde{F}_3=4.890$, there is an alternation between large domains of chaotic trajectories and many very small domains of periodic or unbounded asymptotic solutions (with $\lambda_L<0$) . 
The chaotic trajectories are characterized by positive values of $\lambda_L$ and Poincar\'e sections containing an infinite number of points in the limit of an infinite trajectory.
In the very small domains where $\lambda_L$ becomes negative, the solutions are either asymptotically periodic with, in the Poincar\'e section, a reduced number of points, or unbounded with an absence of points in the x-interval chosen for Fig.~\ref{fig3}a.

At $\tilde{F}_3=4.890$, $\lambda_L$ goes again to 0 and there is a transition towards unbounded asymptotic trajectories ($\lambda_L<0$ and absence of points in the x-interval of the Poincar\'e section Fig.~\ref{fig3}a).
This unbounded domain extends up to the end of the first ID at $\tilde{F}=\tilde{F}_4=5.090$ where again $\lambda_L=0$.

For $\tilde{F}_4<\tilde{F}<\tilde{F}_5$, where $\tilde{F}_5=6.183$ corresponds to the second SD/ID limit, the asymptotic solution of  Eq.~\eqref{eq:maineffect} is again the stable fixed point $(0,0)$ with $\lambda_L<0$, whatever the ICs (sec.~\ref{sssec:converge}).
 
At $\tilde{F}_5$  we enter the second ID and we observe a behavior very similar to the one in the first ID.
The main difference is that the first periodic domain starts directly with two limit cycles, then there is also a cascade of period doublings, the first one at $\tilde{F}=6.522$, leading also to a region of alternation of wide chaotic domains and small periodic or unbounded domains.

The system described by Eq.~\eqref{eq:maineffect} is very rich and complex. The apparition of chaotic solutions is actually not surprising: imposing a standing acoustic wave involves the interaction of two harmonics in a parametric-like excitation.  The  equation~\eqref{eq:maineffect} without dissipation has been studied in order to investigate regimes where resonances induced by both harmonics interact.\cite{Escande1982}
The kicked rotor~\cite{Chirikov1979} is also very similar to our system excepted that an infinity of harmonics are involved in the excitation, so that their resonances can interact leading to chaotic solutions.     
Finally let's mention the work of Van Dooren~\cite{Dooren1996} who studied the dynamics of a pendulum with a forced sinusoidal horizontal support motion: in the absence of gravity, this system reduces to our undamped system ($\gamma=0$). 
\vspace{.5cm}

From this numerical study, it appears that the trajectory of an ad-atom on a substrate submitted to a StAW can be of four different types, depending on the parameters $(\tilde{\gamma},\tilde{F})$. 
While increasing values of $\tilde{F}$ at constant $\tilde{\gamma}$, the domains corresponding to the different kinds of trajectories are successively  i/ converging trajectories in the SD domains, followed in the ID domains by ii/ a periodic domain, then iii/ an alternation of mainly chaotic solutions together with small periodic or unbounded domains and ends up with iv/ an unbounded domain, and so on when entering the next SD.
While the structuring effect of the StAW is obvious in the case of converging trajectories and to a less extend periodic trajectories, since it drives the particle into given regions in the configuration space,  it is less obvious for the other cases, in particular for the chaotic domains.

\section{Stochastic forces}
\label{sec:noise}

In this section, we reintroduce the Gaussian stochastic force  $\xi(t)$  in  Eq.~\eqref{eq:maineffect} in the $\tau_c \to 0$ limit:
\begin{equation}
\ddot{\tilde x}(\tilde{t})+\tilde{\gamma}\dot{\tilde x}(\tilde{t})=\tilde{\xi}(\tilde{t}) + \tilde{F} \cos(\tilde{t}) \sin(\tilde x), \label{eq:stochastic}
\end{equation}
with
\begin{subequations}
\begin{equation}
\left<\tilde{\xi}(\tilde{t})\right>=0, \label{eq:xi_tilde1}
\end{equation}
\begin{equation}
\left<\tilde{\xi}(\tilde{t})\tilde{\xi}(\tilde{t}+\tilde{\tau})\right>=2\tilde{D}\delta(\tilde{\tau}),\label{eq:xi_tilde2}
\end{equation}
\end{subequations}
 where $\tilde{D}=D/(\omega m^2 c_s^2)$ (Eqs.~\eqref{eq:corr-tau} and \eqref{eq:corr-tau1}). 
Note that $\tilde{D}$ and $\tilde \gamma$ are not independent due to the fluctuation-dissipation theorem (Eq~\eqref{eq:flucdis}): 
 \begin{equation} 
 \frac{\tilde{D}}{\tilde{\gamma}} = \frac{k_B T}{m c_s^2} \label{eq:corr2}
 \end{equation}
We will use in this section the same 2.0 fixed value of $\tilde \gamma$ and the same four values of $\tilde F$  as in the previous section, corresponding to the four exhibited types of solutions.
Three values of $\tilde D$ (0.0001, 0.01 and 0.1), or equivalently of temperatures for given substrate ($c_s$) and ad-atom ($m$), will be investigated, corresponding to the quasi-absence of the stochastic force (Sect.~\ref{sec:maineffect}, Eq.~\eqref{eq:maineffect}), a medium and a strong stochastic force respectively.

Due to the stochastic character of  Eq.~\eqref{eq:stochastic} we adopt here a statistical point of view.
It appears, as demonstrated in this section, that on a longtime scale compared to the relaxation time $\tilde{\tau}_R$, the memory of the IC is lost and the system is in a quasi-steady state:   $P(\tilde x)$, the distribution of the ad-atom x-position is mainly time-independent (rigorously, it involves a tiny periodic contribution at the StAW frequency). 
On a timescale larger than the StAW period, the structuring effect of the StAW will be revealed through the correlation between the $\tilde x$ variations of $P(\tilde x)$ and  those of the periodic effective potential $U_{\rm eff}$. Physically, $P(\tilde x)$ will point out the preferential sites where the ad-atom spends most of its time.

Assuming the ergodicity of the system, $P(\tilde x)$ can be obtained  from a single long trajectory of one particle (after elimination of the initial transient period, with whatever IC) or from a set of trajectories. In addition, due to the translational invariance $ \tilde x\to \tilde x + \pi, \tilde t \to \tilde t + \pi $  of Eq.~\eqref{eq:maineffect},  $P(\tilde x)$ is expected to be $\pi$-periodic. 
We calculate $P(\tilde x)$ from a number of trajectories obtained for different ICs and realizations of $\tilde{\xi}(\tilde{t})$: 100 trajectories of  $\tilde{t}=10000$ time units each, 
The ICs are taken at random in the $\tilde x=0$ effective potential valley with $\dot{\tilde x}=0$. The points of the trajectory outside the initial valley are translated back to the $\tilde x=0$ valley coherently with the  translational invariance of Eq.~\eqref{eq:maineffect}. $P(\tilde x)$ is then estimated from the histogram of the ad-atom position of these trajectories and the following normalisation condition:
\begin{equation}
\int_{-\pi/2}^{\pi/2} P(\tilde x) d\tilde x = 1. \label{eq:normalisation}
\end{equation}
The results are reported in Fig.~\ref{fig4} over half the StAW wavelength, i.e. a period of $U_{\rm eff}$ together with the effective potential $U_{\rm eff}$. 
\\

\subsubsection{Converging trajectories}
{\it $\tilde{F}=1.5$, Fig.~\ref{fig4}a.} 
At a very low diffusion  coefficient value ($\tilde{D}=0.0001$), $ P(\tilde x) $ is  strongly peaked at $\tilde x=0$: in the quasi-absence of fluctuating force, the trajectory still converges to the minimum of the effective potential. 
The tiny stochastic force induces small fluctuations of the position in the vicinity of the minimum. 
This fluctuations are not strong enough to induce a transition to an adjacent valley (inter-valley transition) on the simulation time scale ($P(\pm \pi/2)=0$ at the maxima of $U_{\rm eff}$). 
At $\tilde{D}=0.01$, the width of the peak centered on the fixed point of Eq.~\eqref{eq:maineffect} has not sufficiently increased to induce a significant inter-valley transition, whereas it does at $\tilde{D}=0.1$ ($P(\pm \pi/2) \simeq 15\%$): the ad-atom diffuses from a basin of attraction to a neighboring one. The converging character of the trajectory is lost. 
However the self-organization is preserved through the peaked character of $P(\tilde x)$ centered on the minima of $U_{\rm eff}$. 

\subsubsection{ Periodic trajectories}
{\it  $\tilde{F}=3.15$ - Fig.~\ref{fig4}b.}
Here also at very low diffusion  coefficient $\tilde{D}=0.0001$, the periodic character of the trajectory is roughly preserved with small fluctuations around the initial periodic trajectory in the absence of $\tilde \xi$;  and the particle visits a wide region of a basin of attraction of the effective potential.  The peaks and features observed on the plot of $P(\tilde x)$ for $\tilde{D}=0.0001$ are due to the specific shape of the limit cycle (or trajectory shown in Fig.~\ref{fig2}b). Increasing the diffusion coefficient  $\tilde{D}$ induces some fluctuations around this limit cycle. They can even activate the crossing of the effective potential barriers (on the simulation time scale), clearly evidenced for $\tilde{D}=0.1$ by the significant value of $P(\pi/2)$.
 Increasing the diffusion coefficient smooths the structural role of the StAW:  the stochastic fluctuations give  rise to a wide Gaussian-like distribution centered on the effective potential minimum and whose width increases with the diffusion coefficient.  

\subsubsection{Chaotic trajectories}
  {\it   $\tilde{F}=4.5$ - Fig.~\ref{fig4}c.}
In the quasi-absence of stochastic force ($\tilde{D}=0.0001$) the chaotic character of the trajectory leads to a $P(\tilde x)$ distribution correlated to  $U_{\rm eff}$:  even if the ad-atom is not trapped in a given potential valley ($P(\pi/2)\neq 0$ and Fig.~\ref{fig2}c), $P(\tilde x)$ has a pronounced maxima at the minima of $U_{\rm eff}$.
Increasing the diffusion coefficient yields the same qualitative observations as in the previous periodic trajectory case. 
  \subsubsection{Unbounded trajectories}
\begin{figure}
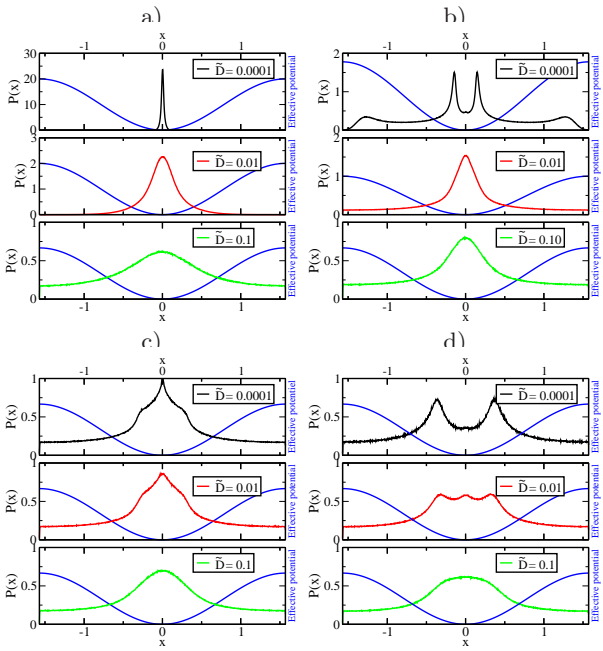

    \centering
    \begin{tabular}{cc}
    a) & b) \\
      \includegraphics[width=0.45\columnwidth]{fig4a}
       &  \includegraphics[width=0.45\columnwidth]{fig4b}\\
      c) & d) \\
      \includegraphics[width=0.45\columnwidth]{fig4c}
      &  \includegraphics[width=0.45\columnwidth]{fig4d}\\
	\end{tabular}
    \caption{Black, red and green curves : Histograms of $P(\tilde x)$ for different values of the diffusion coefficient $\tilde{D}$: 0.0001, 0.01 and 0.1 respectively, with $\tilde{\gamma}=2.0$ and different $\tilde{F}$ values, i.e. types of trajectories as defined in sec.~\ref{ssec:num_study}: (a) converging ($\tilde{F}=1.5$), (b) periodic ($\tilde{F}=3.15$), (c) chaotic ($\tilde{F}=4.5$) and (d) unbounded ($\tilde{F}=4.96$). Blue curve: $U_{\rm eff}(X)$, the effective potential in arbitrary units.}
    \label{fig4}
\end{figure}

   {\it  $\tilde{F}=4.96$ - Fig.~\ref{fig4}d.}
At the very low $\tilde{D}=0.0001$ value, from the staircase character of the trajectory presented in Fig.~\ref{fig2}d  one expects a non uniform $P(\tilde x)$ with a marked peak inside each potential well.
A precise examination of the trajectory in Fig.~\ref{fig2}d reveals that the plateaux are at $\tilde x$ positions slightly larger than the center of the potential wells.
Such a trajectory contributes then to the $\tilde x_1 >0$ peak observed on Fig.~\ref{fig4}d (top). The second peak at the symmetric $-\tilde x_1$ position results from mean trajectories towards $\tilde x <0$  positions.
The $\tilde x_1 >0$ non centered position of the distribution associated to trajectories towards increasing mean $\tilde x$  values results from the definite direction $\tilde x>0$ or $\tilde x<0$ of the observed trajectories ($\tilde x>0$ in Fig.~\ref{fig2}d).
Increasing the diffusion coefficient results in an increased width of each peak, leading to a unique central peak for $\tilde{D}=0.1$.
As in the three preceding cases, the structuring effect of the StAW is also evidenced in that case whatever the strength of the stochastic force in the studied range.

As a  conclusion, we would like to emphasize that for all the trajectories types, and as soon as the stochastic force is significant, the distribution $P(\tilde x)$ is a  Gaussian-like distribution centered on the effective potential minimum. Therefore, the StAW has a structuring effect on the diffusion of the particle: it induces preferential sites in which the particle spends more time. These sites precisely correspond to the minima of the effective potential evidenced in Sect.~\ref{sec:maineffect}. The stochastic force essentially counterbalances the structurating role of the StAW by smoothing the distribution $P(\tilde x)$.

\section{Retarded effects}
\label{sec:retarded}

We have neglected the correlation time $\tau_c$ in Sects.~\ref{sec:maineffect} and~\ref{sec:noise}. In surface diffusion problems, such an approximation is valid for heavy ad-atoms\cite{Forster1990} which is a very peculiar case.
  In this section, we investigate the effect of a significant value of the correlation time $\tau_c$ compared to the StAW period.  The correlation time $\tau_c$ is involved in the friction term through the memory kernel (Eq.~\eqref{eq:mem_kernel}) and in the auto-correlation function of the stochastic force $\xi(\tilde t)$ (Eq.~\eqref{eq:corr}).  
Reintroducing the retarded effects  Eqs.~\eqref{eq:stochastic},  \eqref{eq:xi_tilde1} and \eqref{eq:xi_tilde2} write now:
\begin{eqnarray}
\ddot{\tilde x}(\tilde{t})+\frac{\tilde{\gamma}}{\tilde{\tau}_c}\int_{-\infty}^{\tilde{t}} \dot{\tilde x}(\tilde{t}')e^{-(\tilde{t}-\tilde{t}')/\tilde{\tau}_c } {\rm d\tilde{t}'}
=\tilde{F} \cos(\tilde{t}) \sin(\tilde x) + \tilde{\xi}(\tilde{t}), \nonumber\\
\label{eq:retarded}
\end{eqnarray}
\begin{subequations}
\begin{equation}
\left<\tilde{\xi}(\tilde{t})\right>=0, \label{eq:mean_xi2}
\end{equation}
\begin{equation}
\left<\tilde{\xi}(\tilde{t})\tilde{\xi}(\tilde{t}+\tilde{\tau})\right>=\tilde{D} \frac{ e^{-|\tilde{\tau}|/\tilde{\tau}_c} }{\tilde{\tau}_c }, \label{eq:correl_xi2}
\end{equation}
\end{subequations}
with the same Eq.~\eqref{eq:corr2} between $\tilde{D}$ and $\tilde{\gamma}$.

Let's first estimate a physical range for the time $\tilde{\tau}_c= \omega \tau_c$.  On the one hand, in the case of the diffusion of an ad-atom, $1/\tau_c$ is roughly of the order of the Debye frequency $f_{D}$, corresponding to the maximum frequency of  atomic vibrations in the crystalline substrate, i.e. $f_{D} \approx 10^{13}$ Hz for common crystals, and consequently  $\tau_c \approx 0.1$ ps. 
On the other hand,  as mentioned in the introduction,  the StAW wavelengths of interest vary  from few to hundreds of nanometers, i.e. typically from $5$ nm to  $1$ $\mu$m. 
With typical sound speeds in solids around $3000$ m.s$^{-1}$, the StAW frequency $\omega$ lies in the range $1.25$ 10$^{10}$  - $2.5$ 10$^{12}$ rad.s$^{-1}$.
Consequently, $\tilde{\tau}_c$ will be in the range 0.002 - 0.40. 
 
 Three typical values of $\tilde{\tau}_c$, 0.001, 0.3 and 0.4 will be used to investigate the effect of $\tau_c$ in the two extreme cases: quasi absence  ($\tilde D=0.0001$) and significant ($\tilde D=0.01$) fluctuations, two values of $\tilde D$ already used in the preceding sections.
Due to the separation of the different time scales, we expect the solutions of Eq.~\eqref{eq:retarded} with a low $\tilde{\tau}_c$ value (0.001) to be very similar to solutions of Eq.~\eqref{eq:stochastic}. 
Only values of $\tilde{\tau}_c$ non negligible compared to the StAW period are expected to produce solutions of  Eq.~\eqref{eq:retarded} significantly different from those of  Eq.~\eqref{eq:stochastic}.
We reduce our study to the converging trajectory case previously studied $(\tilde{\gamma} = 2, \tilde{F}=1.5)$, with the same (1.0,0.0) IC as in Sect.~\ref{sssec:converge},  since this type of trajectory is the most favorable for self-organisation and thus will be usually preferred in any application.

Eq.~\eqref{eq:retarded} is numerically solved using a Leap-frog algorithm.\cite{frenkel} 
$\tilde{\xi}(\tilde{t})$ values satisfying Eqs.~\eqref{eq:mean_xi2} and ~\eqref{eq:correl_xi2} are generated with the algorithm of Ref.~\onlinecite{Fox1988} while the integral of the friction term including the memory kernel are calculated using the algorithm given in  Ref.~\onlinecite{Gordon2008}.

\begin{figure}
\includegraphics[width=0.9\columnwidth]{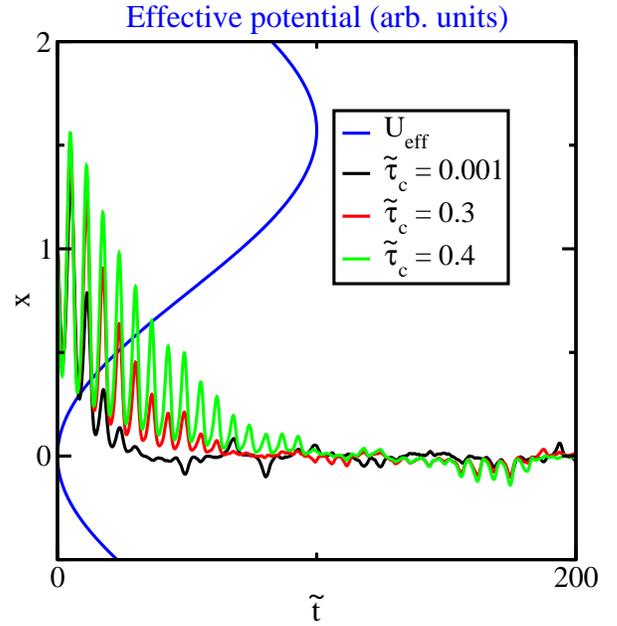}         
  \caption{Converging trajectories $\tilde x (\tilde t)$ solutions of Eq.~\eqref{eq:retarded}, $(\tilde{\gamma},\tilde{F})=(2.0,1.5)$, with a small diffusion coefficient ($D=0.0001$) and initial conditions $(\tilde x,\dot{\tilde x})=(1.0,0.0)$, for three different values of $\tilde{\tau}_c$:  0.001 (black), 0.3 (red), 0.4 (green). Blue curve : $U_{\rm eff}(X)$, the effective potential  (abscissa) in arbitrary units versus position  $X$ (ordinate).} 
\label{fig5}
\end{figure}

\begin{figure}
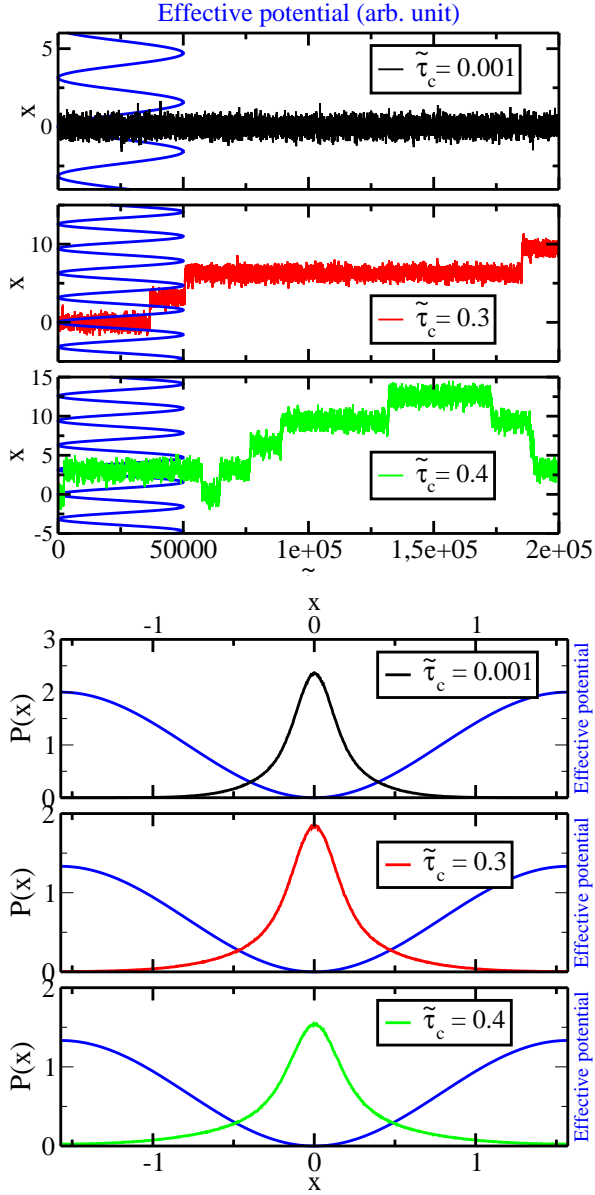

    \centering
\includegraphics[width=0.9\columnwidth]{fig6a}        
 \includegraphics[width=0.9\columnwidth]{fig6b}        
    \caption{(a) : converging trajectories $\tilde x (\tilde t)$ solutions of Eq.~\eqref{eq:retarded}, $(\tilde{\gamma},\tilde{F})=(2.0,1.5)$, with a diffusion coefficient $D=0.01$ and initial conditions $(\tilde x,\dot{\tilde x})=(1.0,0.0)$, for three different values of $\tilde{\tau}_c$:  0.001 (black), 0.3 (red), 0.4 (green). Blue curve : $U_{\rm eff}(X)$, the effective potential  (abscissa) in arbitrary units versus position  $X$ (ordinate).
(b): Corresponding histograms of $P(\tilde x)$ from 100 trajectories of 10 000 time units with initial conditions evenly distributed between $-\pi$ and $\pi$ (same color code as in (a)).}
    \label{fig6}
\end{figure}

The solutions for $\tilde{D}=0.0001$ are reported on Fig.~\ref{fig5}.
As expected, when the fluctuations are negligible, whatever the correlation time, the  solutions of Eq.~\eqref{eq:maineffect} are qualitatively unchanged and thus  $\tau_c$ has a negligible influence on the structuring effect.
The solutions are still oscillating functions at the StAW time scale and they still converge to the fixed point. 
The main effect of increasing values of $\tau_c$ is an increase of the amplitude of the oscillations at all timescales, and correlatively of the relaxation time $\tilde{\tau}_R$ of the ad-atom dynamics.
Such behavior has already been mentioned and explained in the literature: the velocity autocorrelation function for the ad-atom shows both a ballistic and a diffusive regime,  the width, roughly $\tilde{\tau}_R$ of the balistic regime, is related to $\tilde{\tau}_c$.\cite{Gordon2008}

At odds, when the stochastic force is large ($\tilde{D}=0.01$), the possibility for the ad-atom to cross an effective potential barrier on the simulation time scale increases with $\tilde{\tau}_c$: we report in Fig.~\ref{fig6}a the trajectories of the ad-atom as a function of time for the three different mentioned values of $\tilde{\tau}_c$. Increasing $\tilde{\tau}_c$, and hence the ballistic regime,  gives more importance to the very unlikely extreme values of $\tilde{\xi(t)}$ and thus results in a higher effective diffusion coefficient (not shown).  
Fig.~\ref{fig6}b reports the $P(\tilde x)$ distributions:  $\tilde{\tau}_c$ slightly affects the histogram $P$ in the vicinity of the minimum of $U_{\rm eff}$. The effect is more pronounced on its wings: they increase significantly leading to non zero $P$ values at the $U_{\rm eff}$ maxima for $\tilde{\tau}_c=0.3$ and $0.4$, coherently with the possibility for the ad-atom to escape from its original potential valley evidenced in Fig.~\ref{fig6}a).
Nevertheless, no matter the value of  $\tilde{\tau}_c$, the shape of $P(\tilde x)$ is still gaussian-like evidencing the structuring effect of the StAW

We can thus conclude that if the retarded effects quantitatively modify the trajectories, they weakly affect the structuring effect induced by the StAW. 


\section{Effective Crystalline potential}
\label{sec:crystal}

In this section, we consider the additional effect of the effective crystalline potential $\Phi_{\rm eff}(\tilde x)$ on the motion of the ad-atom, in the same  negligible correlation time  limit ($\tau_c \to 0$).  Eq.~\eqref{eq:totalreduced} writes then:
\begin{eqnarray}
\ddot{\tilde x}(\tilde{t})&+&\tilde{\gamma} \dot{\tilde x}(\tilde{t})= -\tilde{\varphi_0} \sin(n \tilde x) 
+ \tilde{F} \cos(\tilde{t}) \sin(\tilde x)+\tilde{\xi}(\tilde{t}) \nonumber \\
 \label{eq:poteff}
\end{eqnarray}

The calculations will be performed with $n=24$, a value comparable to the ones we used in our MD simulations.\cite{Taillan2011}

\begin{figure}
\includegraphics[width=0.9\columnwidth]{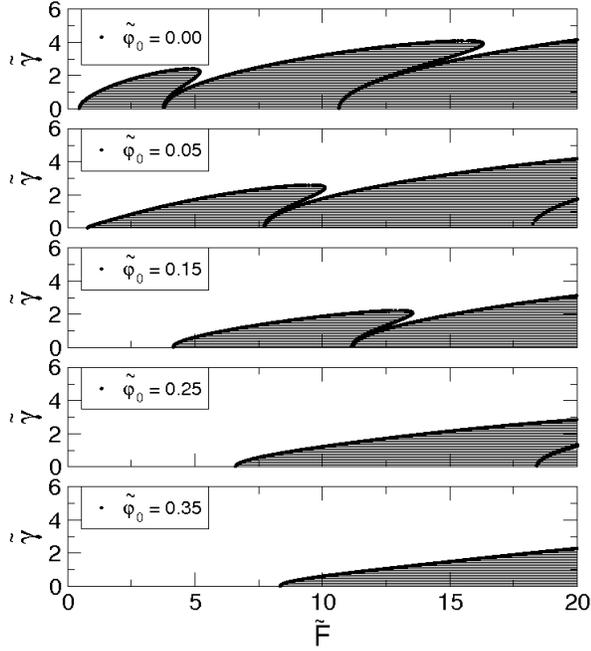}         
  \caption{Stability (blank) and instability (dashed) domains of the fixed point (0,0) of Eq.~\eqref{eq:poteff}  in the ($\tilde{\gamma}$,$\tilde{F}$) parameter plane (PP) for increasing values of the  effective cristalline force $\tilde{\varphi_0}$. From top to bottom : $\tilde{\varphi_0}=0.00$ (same as Fig.~\ref{fig1}), 0.05, 0.15, 0.25 0.35.}
\label{fig7}
\end{figure}

First, we study the modifications induced by $\Phi_{\rm eff}$ on the fixed-point stability diagram described in Sect.~\ref{sec:maineffect} - Fig.~\ref{fig1} in absence of stochastic forces. 
Note that in Eq.~\eqref{eq:poteff} both the crystalline and StAW forces cancel for $\tilde x=0$.
We could easily imagine that a dephasing of $\Phi_{\rm eff}$ compared to the StAW will shift the fixed points or yield to the absence of fixed points. However, since the StAW wavelength is large compared to the lattice parameter, we do not expect such a dephasing to qualitatively modify the ad-atom trajectories, especially in the presence of the stochastic force.   
Fig.~\ref{fig7} reports, as in Fig.~\ref{fig1}, the stability diagram in the ($\tilde{\gamma}$,$\tilde{F}$) PP  for increasing values of $\tilde{\varphi_0} = 0.0, 0.05 , 0.15, 0.25$ and $0.35$.
The main effect of $\Phi_{\rm eff}$ is to shift the instability domains towards higher $\tilde{F}$ values at constant $\tilde{\gamma}$. 
The effective crystalline potential has then a stabilizing effect on the trajectories.  


 To investigate the structuring effect of the StAW, as in the preceding section, we will study the effect of $\Phi_{\rm eff}$ in the presence of the stochastic force only for one couple  $(\tilde{\gamma} = 2, \tilde{F}=1.5)$, corresponding to converging trajectories in the absence of crystalline potential. 
Here we are mainly interested on what happens when $\Phi_{\rm eff}$ is substantially higher than the effective StAW potential $U_{\rm eff}$ and is thus potentially able to challenge or to overcome the structuring effect of the StAW: actually, the StAW force becomes then a second order effect compared to the crystalline potential. 
In order to induce an efficient diffusion across the effective crystalline potential barrier on the simulation time scale, we will use the strongest stochastic force ($\tilde{D}=0.1$) previously used.

\begin{figure}
    \centering
 \includegraphics[width=.9\linewidth]{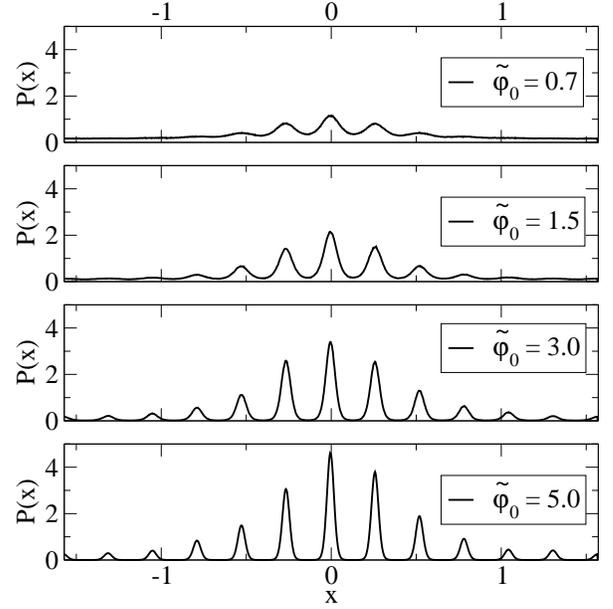}\\
    \caption{Histogram of $P(\tilde x)$ calculated from  converging trajectories ($(\tilde{\gamma},\tilde{F})=(2.0,1.5)$) solutions of Eq.~\eqref{eq:poteff}  ($100$ trajectories of $200000$ time units each), with $\tilde{D}=0.1$  and increasing values of $\tilde{\varphi_0}$  from top to bottom: $\tilde{\varphi_0}=0.7, 1.5, 3, 5$. 
    }
    \label{fig8}
\end{figure}

Solving numerically Eq.~\eqref{eq:poteff}, we calculate the histogram $P(\tilde x)$  from 100 trajectories of 50000 time units each with different ICs.
Fig.~\ref{fig8} reports the  distribution $P(\tilde x)$ (normalized following Eq.~\eqref{eq:normalisation}) as a function of $\tilde x$ for increasing values of $\tilde{\varphi_0}=0.7, 1.5 , 3.0$ and $5.0$, starting at a lower value than $\tilde F$ up to 3.3  times $\tilde F$.
Comparing to Fig~\ref{fig6}, we see that the Gaussian like curves present now a structuration at the $\Phi_{\rm eff}$ length scale.
There are now two length scales: 
1) a short wavelength oscillation (lenght scale $a$) due to $\Phi_{\rm eff}$ with the local maxima of $P(\tilde x)$ at the minima of $\Phi_{\rm eff}$. 
2) a slow variation of the amplitude of the local maxima corresponding to the previous curves of Fig~\ref{fig6} and thus to the StAW with the maxima of $P(\tilde x)$ at the minima of $U_{\rm eff}$.
Unexpectedly, from Fig.~\ref{fig8},  the structuring effect seems to increase with the increasing of the effective crystalline potential. However, this effect is essentially due to the normalization: fitting the set of maxima of $P(\tilde x)$ by a Gaussian curve leads to approximately the same Gaussian width for all the values of $\tilde{\varphi_0}$.

Hence, the structuring effect is weakly affect by the values (even for significant values) of $\tilde{\varphi_0}$. 
Note the very good qualitative agreement between  the distribution $P(\tilde x)$ in Fig.~\ref{fig8} and the results reported in our Molecular dynamics simulations~\cite{Taillan2011}. 

\section{Effective potential and probability distribution}
 \label{sec:pot_eff_distrib}
 
\begin{figure}
\includegraphics[width=0.9\columnwidth]{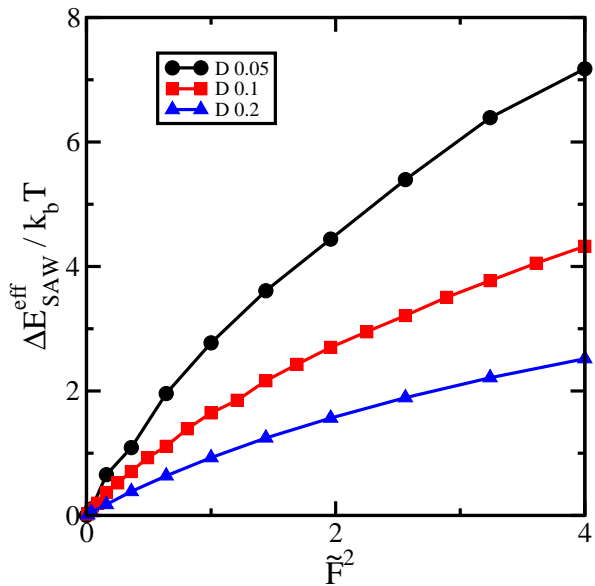}         
  \caption{Effective energy difference $\Delta {E}^{\text{eff}}_{\text{SAW}}$ as a function of ratio $\tilde{\varphi_0}/\tilde{F}$ calculated from histograms $P(\tilde x)$.  Histogram $P(\tilde x)$ are calculated from $100$ trajectories of $200000$ time units for each values of $\tilde{\varphi_0}/\tilde{F}$ by solving Eq.~\eqref{eq:poteff} with $\tilde{\gamma}=2.0$, $\tilde{\varphi_0} = 3.0$, $\tilde{D}=0.001,0.01,0.1$.} 
\label{fig9}
\end{figure}

 
To quantify the structuring effect of the StAW, we note that the structuration can be characterized by an effective energy difference $\Delta {E}^{\text{eff}}_{\text{SAW}}$ at the mesoscopic scale between the minima and maxima of the effective potential deduced from the $P(\tilde x)$ curves of Fig.~\ref{fig8}:
\begin{equation}
\Delta {E}^{\text{eff}}_{\text{SAW}} = k_BT \ln \left[\text{max}(P_{\text{max}})/\text{min}(P_{\text{max}})\right] \label{Eeff}
\end{equation}
 with $P_{\text{max}}$ the ensemble of local maxima of $P(\tilde x)$ and max (min) the ensemble maximum(minimum). 
Since both effective energy differences, $\Delta {E}^{\text{eff}}_{\text{SAW}}$ (Eq.~\eqref{Eeff}) and effective potential $U_{\rm eff}$ (Eq.~\eqref{ueff}), govern the diffusion of the ad-atom, and since   $U_{\rm eff}$  quadratically depends on the amplitude of the force $\tilde{F}$, we can reasonably expect the same quadratic dependence for $\Delta {E}^{\text{eff}}_{\text{SAW}}$. However, note that while  $U_{\rm eff}$ does not take into account the stochastic force $\tilde{\xi}(t)$ nor the effective crystalline potential, $\Delta {E}^{\text{eff}}_{\text{SAW}}$  implicitly takes into account all these contributions.  Fig.\ref{fig9} exhibits the dependence of $\Delta {E}^{\text{eff}}_{\text{SAW}}$ as a function of $\tilde{F}^2$ for $\tilde{\gamma}=2$, $\tilde{\varphi_0} = 3.0$ and for $D=0.05$, $0.1$ and $0.2$.   $\Delta {E}^{\text{eff}}_{\text{SAW}}$ has been calculated from $P(\tilde x)$ curves similar to the ones of Fig.~\ref{fig8}.  We did not succeed to calculate $\Delta {E}^{\text{eff}}_{\text{SAW}}$ as a function of $\tilde{F}$ for $D<0.05$:  the average time needed by the particle to escape from an effective crystalline potential being too long to obtain good statistics, due to the activated character of this event with a $1/D$ exponential dependence.   
The linear dependence of $\Delta {E}^{\text{eff}}_{\text{SAW}}$  as a function of $\tilde{F}^2$ is observed in the small forces regions where the perturbative calculations of Sect.~\ref{sec:multiple} scientifically sound.  We note that increasing the diffusion coefficient or the friction $\tilde{\gamma}$ (not shown) yields a smaller StAW structuring effect i.e. as already mentioned, the thermal noise challenges the StAW structuring effect.

 The results of this section demonstrates  that  the StAW amplitude and the temperature, acting on the friction $\gamma$ and the diffusion coefficient $D$ allow to tune the structuration induced by the StAW.  These information are essential to identify the key parameters if one wishes to experimentally implement the dynamic substrate structuring effect described here.

\section{Conclusion}

In this work,  we have studied the solutions of the motion equation of an ad-atom diffusing on a substrate submitted to a StAW. To this aim, we have applied some standard tools of the non-linear physics to a simplified version of the motion equation keeping the most relevant terms. Noticeably, an effective potential governing the slow dynamics of the ad-atom has been derived. We have shown that this effective potential controls the distribution of the ad-atom x-position even when considering  the other additional terms (noise, retarded effects, effective crystalline potential) in the ad-atom motion equation. 
 We underline the relevance of our model when comparing calculated distribution probabilities to positions histograms collected from molecular dynamics simulations~\cite{Taillan2011}.

Our present study opens different perspectives.  \\
In a recent publication~\cite{Taillan2011}, we have discussed about the possibilities to experimentally implement the  dynamic substrate structuring effect. We have noticeably mentioned  the impossibility to produce StAW with a wavelength in the nanometer range on the substrate surface using the current available experimental setups: to our knowledge, and in the state of the art, the production of surface acoustic waves with wavelengths in the 100 nm range is possible using optical excitations~\cite{siemens2009}. However, the dynamic structurating effect does not directly depend on the wavelength. Actually, the dynamic effect is expected to exist as long as the ad-atom does not have the time to diffuse on a wavelength $\lambda$ during a period of the StAW  i.e. 
\begin{equation} 
k \ll \frac{c_s}{D_{\rm eff}}
\end{equation} 
where $D_{\rm eff}$ is the effective diffusion coefficient of the ad-atom in the effective crystalline potential. 
 One may thus consider the possibility to use StAW with hundred nanometers wavelengths. Our present model and study will then be a very fruitful tool to evince the optimized parameters (especially the temperature and the StAW amplitude) leading to an efficient structuring effect. 

Theoretically,  Eq.~\eqref{eq:maineffect} exhibits a very rich and complex dynamics. We have studied the solutions of this equation, but we currently consider the possibility to derive the Fokker-Planck equations for the ad-atom probability density functions.  Such equations would allow to directly derive the position distribution $P(\tilde x)$, and even perhaps to derive the effective diffusion coefficient describing the diffusion of the ad-atom between the different basins of the effective potential.  
 
 Finally, in this paper, we have focus  on the structuring effect induced by the StAW on the position distribution of the ad-atom and have eluded the study of the  dynamics of the ad-atom. We have nevertheless exhibited  the possibility for the ad-atom to follow different qualitative types of trajectories depending on the parameters $\tilde{\gamma}$, $\tilde{F}$ and $\tilde{\varphi_0}$.  The study of the ad-atom dynamics, including the dependence of the StAW force on the substrate lattice parameter scale is a natural perspective to this work. 
 

%
\end{document}